\documentclass[aps,prb,notitlepage,twocolumn]{revtex4-1}

\usepackage{amsmath}
\usepackage{amssymb}
\usepackage{graphicx}
\usepackage{epstopdf}
\usepackage[colorlinks=true]{hyperref}

\newcommand{\sech}{\mathrm{sech}}

\begin{document}
\title{The dynamics of bimeron skyrmions in easy-plane magnets induced by a spin supercurrent}

\author{Se Kwon Kim}
\affiliation{Department of Physics and Astronomy, University of Missouri, Columbia, Missouri 65211, USA}
\date{\today}

\begin{abstract}
We theoretically study the interaction of an isolated bimeron skyrmion in quasi-two-dimensional easy-plane magnets with a surrounding spin superfluid associated with spontaneously broken U(1) spin-rotational symmetry, revealing that skyrmion energy depends on the local spin current flowing in its background. The finding leads us to propose to manipulate a skyrmion energy landscape via a spin supercurrent, which can be controlled non-locally by varying the magnitudes of spin-current injection and ejection through the boundaries. Two exemplary cases are discussed: a steady-state motion of a skyrmion induced by a uniform force and a skyrmion motion localized along a one-dimensional racetrack. We envision that a skyrmion interacting with a spin superfluid can serve as a robust point-like information carrier that can be operated with minimal dissipation. 
\end{abstract}

\maketitle

\section{Introduction}
\begin{figure}
\includegraphics[width=0.8\columnwidth]{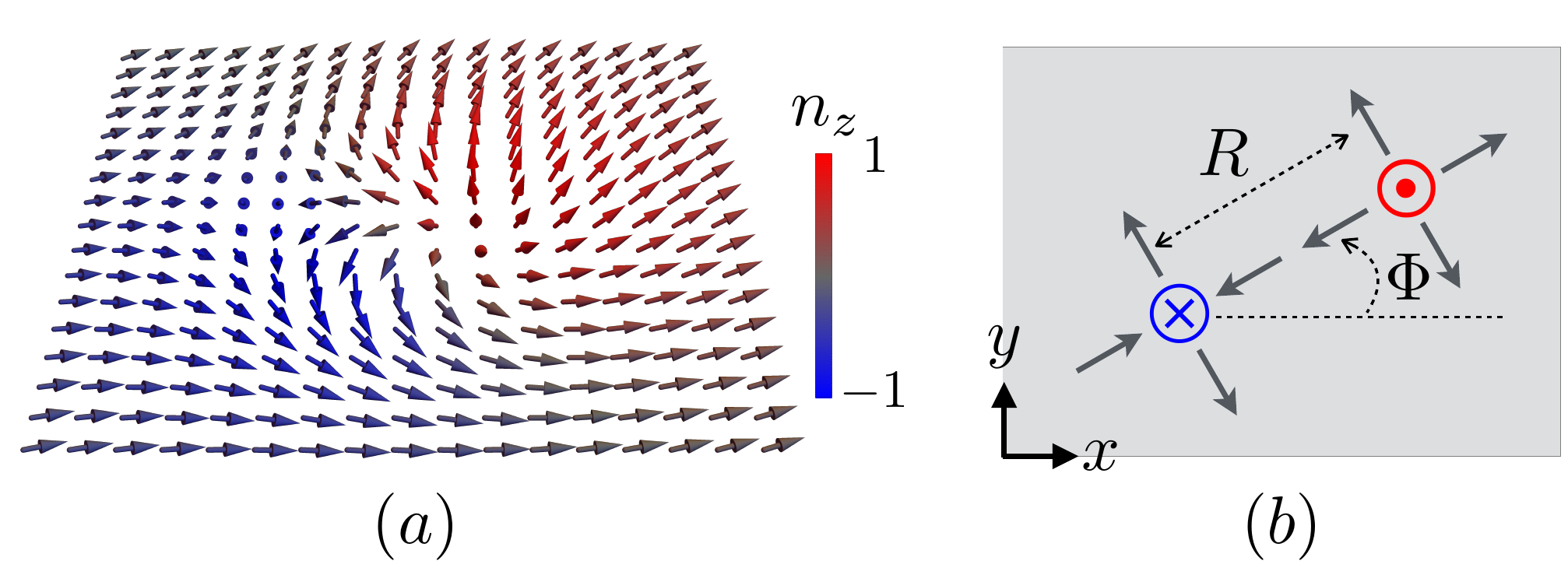}
\caption{(a) Spin configuration of a bimeron skyrmion in an easy-plane magnet. (b) Illustration of a skyrmion as a composite object of a vortex with spin-up core and an antivortex with spin-down core, which are also referred to as merons.}
\label{fig:fig1}
\end{figure}

Topological solitons in magnets have been attracting great attention for many decades~\cite{*[][{, and references therein.}] KosevichPR1990}. One example is a skyrmion, a swirling spin texture in two dimensions, which has been of surging interest during the last decade due to its promising role as a robust point-like information carrier in spintronic devices~\cite{*[][{, and references therein.}] FertNN2013, *[][{, and references therein.}] NagaosaNN2013}. The previous theoretical and experimental efforts have been mainly focused on skyrmions in easy-axis chiral magnets, which are stabilized by certain spin-orbit coupling~\cite{BogdanovJETP1989, MuhlbauerScience2009, YuNature2010}. There have been two recent developments in searching for new platforms for skyrmions. First, theoretical and numerical investigations have shown that skyrmions can be stabilized by frustrated Heisenberg exchange interactions even in the absence of any spin-orbit coupling~\cite{OkuboPRL2012, HayamiPRB2016, LinPRB2016, LeonovPRB2017, LeonovNC2017, LiangNJP2018}. The exchange interactions respect spin-rotational symmetry unlike spin-orbit interactions, and thus the resultant skyrmions can possess extra degrees of freedom associated with the symmetry compared to the conventional ones~\cite{LeonovNC2015}. Second, magnets with easy-plane anisotropy have been emerging as an alternative material platform for skyrmions~\cite{WaldnerJMMM2008, JaykkaPRD2010, BanerjeePRX2014, MoonarXiv2018, MurookaarXiv2018}. The skyrmions thereof can be viewed as composite objects of two magnetic vortices also known as merons~\cite{KharkovPRL2017, GobelPRB2019} (see Fig.~\ref{fig:fig1}). \textcite{YuNature2018} reported the observation of transformation between such merons and skyrmions in an easy-plane magnet. 

Easy-plane magnets with spin-rotational symmetry have a zero-energy mode associated with the spontaneously broken U(1) symmetry. In 1978, \textcite{SoninJETP1978} showed theoretically that such systems can support superfluid-like spin transport analogous to superfluid mass transport in Helium-4 where the U(1)-phase symmetry is spontaneously broken. The interest on superfluid spin transport has been revived recently by advancements in spintronic techniques for a spin current, gathering significant attention owing to its ability for long-distance low-dissipation spin transport~\cite{KonigPRL2001, SoninAP2010, ChenPRB2014, TakeiPRL2014, ChengPRB2014, ChenPRL2015, IacoccaPRB2017, QaiumzadehPRL2017}. In particular, \textcite{TakeiPRL2014} showed that superfluid spin transport can be realized in magnets by injecting and ejecting a pure spin current through their interfaces with normal metals via the spin Hall effect. The resultant spin supercurrent decays algebraically in space differing from an exponentially decaying diffusive spin current, as seen in two spin-transport experiments in insulators, Cr$_2$O$_3$~\cite{YuanSA2018} and quantum Hall graphene~\cite{StepanovNP2018}. Previously, the spin supercurrent has been shown to be able to induce a motion of a domain wall, which is a topological solitons in magnets with spontaneously broken discrete symmetry, both in easy-cone magnets~\cite{KimPRB2016-2} and in bilayers of easy-axis and easy-plane magnets~\cite{UpadhyayaPRL2017}.

In this work, we explore the possibility to control a skyrmion in easy-plane magnets non-locally via a spin superfluid that is controlled through the boundary. To this end, we theoretically study the interaction of a skyrmion with a spin supercurrent flowing in its background. The skyrmion energy is shown to depend on the magnitude of the local spin supercurrent, which enables us to engineer the energy landscape by an inhomogeneous spin supercurrent and thereby drive a skyrmion. We provide two examples of such control: a steady-state motion of a skyrmion driven by a uniform force and a realization of a skyrmion racetrack. One promising material candidate is offered by frustrated triangular magnets with easy-plane anisotropy such as NiBr$_2$\cite{DayJPC1976, LeonovNC2015, KharkovPRL2017}, whose exchange interactions can be tuned by chemical substitutions. We envision that a bimeron skyrmions and a spin superfluid in easy-plane magnets can serve as the pair of a point-like memory unit and an efficient controller.

The paper is organized as follows. In Sec.~\ref{sec:2}, we introduce a bimeron skyrmion that can be stabilized in easy-plane frustrated magnets. In Sec.~\ref{sec:3}, we study its interaction with a spin supercurrent that is carried by a spiraling spin texture in easy plane. In particular, we discuss a one-dimensional motion of a bimeron skyrmion induced by a spin supercurrent with a uniform gradient in Sec.~\ref{sec:3-1} and proposal for realizing a skyrmion racetrack by using a spin supercurrent in Sec.~\ref{sec:3-2}. We will conclude the paper in Sec.~\ref{sec:4} by discussing certain approximations made in our discussions and providing a future outlook.

\section{Bimeron skyrmion in frustrated magnets}
\label{sec:2}
Our model system is a quasi-two-dimensional easy-plane magnet whose static energy can be described by the following energy functional:
\begin{equation}
\label{eq:U}
U = \frac{1}{2} \int dxdy \left[ A (\nabla \mathbf{n})^2 + K n_z^2 + B (\nabla^2 \mathbf{n})^2 \right] \, , 
\end{equation}
where $\mathbf{n} = (n_x, n_y, n_z)$ is the unit vector in the direction of the magnetic order (e.g., magnetization in ferromagnets and Neel order in antiferromagnets) and $A, B$, and $K$ are positive parameters. The first two terms are the quadratic exchange and the easy-plane anisotropy energies, respectively, which are in the conventional miromagnetic treatment of magnetic energy. The last term is a quartic exchange term, which has been shown to exist in certain magnets with frustrated exchange interactions~\cite{LeonovNC2015, HayamiPRB2016, KharkovPRL2017}. It is well known that a soliton with a finite size cannot be stabilized by the first two terms only, which can be understood by invoking the Hobart-Derrick's scaling argument~\cite{HobartPPS1963, *DerrickJMP1964}: The energy of such a texture can be lowered by shrinking its size uniformly, $\mathbf{n} (\mathbf{r}) \mapsto \mathbf{n} (\mathbf{r} / \lambda)$ with $\lambda < 1$ and thus it is unstable. The quartic term can stabilize a soliton by penalizing shrinking. In particular, the Hamiltonian has been shown to support a skyrmion in the form of a pair of two merons, vortex and antivortex, by analytical and numerical calculations in Refs.~\cite{KharkovPRL2017, LeonovNC2017, GobelPRB2019}.

To describe a bimeron skyrmion with skyrmion number $Q = \pm 1$ defined by
\begin{equation}
Q = \frac{1}{4 \pi} \int dxdy \, \mathbf{n} \cdot (\partial_x \mathbf{n} \times \partial_y \mathbf{n}) \, ,
\end{equation}
we adopt the variational ansatz used in Refs.~\cite{KharkovPRL2017, GobelPRB2019}:
\begin{equation}
\label{eq:ansatz}
\mathbf{n} = \mathcal{R}_{\hat{\mathbf{z}}} (\phi) \mathcal{R}_{\hat{\mathbf{x}}} (Q \Phi) \mathcal{R}_{\hat{\mathbf{y}}} (- \pi Q/2) \, \mathbf{n}_0 \, ,
\end{equation}
where $\mathcal{R}_{\hat{\mathbf{a}}} (\varphi) $ is the rotation matrix by angle $\varphi$ with respect to the axis $\hat{\mathbf{a}}$, $\mathbf{n}_0 = (\sin \zeta_0 \cos \eta_0, \sin \zeta_0 \sin \eta_0, \cos \zeta_0)$ is the ansatz for a skyrmion in an easy-axis magnet: $\zeta_0 = (1 + Q) \pi / 2 - \pi Q \exp(- r / R)$ with $\eta_0 = \arctan[(y - Y) / (x - X)]$, and $r = [(x - X)^2 + (y - Y)^2]^{1/2}$ is the distance from the skyrmion center. The skyrmion is described by four parameters $X, Y, \Phi$ and $R$: $X$ and $Y$ are the skyrmion positions, which represent zero-energy modes associated with the translational invariance of the system; $\Phi$ is the angle from the antivortex to the vortex, which is arbitrary and thus represents another zero-energy mode; $R$ represents the skyrmion size, or, equivalently, the distance between two constituent merons. Its equilibrium value is determined by competition of the anisotropy and the quartic exchange energies~\cite{KharkovPRL2017}: $R_0 = C (B/K)^{1/4}$, where $C$ is a dimensionless number of order of 1. The bimeron skyrmion with this equilibrium size has been shown analytically and numerically to be stable within the current model [Eq.~(\ref{eq:U})] in Ref.~\cite{KharkovPRL2017}. The variable $\phi$ is the azimuthal angle of the background, which represents the zero-mode associated with the spontaneously broken U(1) spin-rotational symmetry. Figure~\ref{fig:fig1}(a) and (b) show the skyrmion spin textures with $\Phi = \pi/6$ and $\phi = 0$ and the schematic for its composition, respectively. From now on, we will assume that a bimeron skyrmion is already present in the system.

\section{Interaction of a bimeron skyrmion with a spin supercurrent}
\label{sec:3}
Sufficiently far from a skyrmion, the order parameter stays closely within the easy plane and thus the order parameter configuration can be well described by its azimuthal angle $\phi$. Since the azimuthal angle $\phi$ and the spin density $s_z$ along the z axis form a pair of canonical conjugate variables, the long-wavelength dynamics can be captured by $U \approx \int dxdy \, [A (\boldsymbol{\nabla} \phi)^2 / 2 + s_z^2 / (2 \chi)]$, where $\chi$ is the magnetic susceptibility. For ferromagnets, the magnetic susceptibility is given by $\chi = s^2 / K$ [see Eq.~(\ref{eq:U})], where $s$ is the saturated spin density. For antiferromagnets, the term $\propto s_z^2$ is not included in the static energy functional given in Eq.~(\ref{eq:U}) because it is strongly suppressed by the antiferromagnetic exchange in static cases, but it needs to be included to describe the dynamics of the azimuthal angle $\phi$. Within linear response, the dynamics of both easy-plane ferromagnets and antiferromagnets can be described in terms of the same pair of canonical conjugate variables, $\phi$ and $s_z$, which allows us to treat a spin current in the both systems in a single framework as discussed in Ref.~\cite{HalperinPR1969, TakeiPRL2014, TakeiPRB2014}. See Apps.~\ref{app:fm} and~\ref{app:afm} for the detailed discussions about the low-energy dynamics of easy-plane ferromagnets and antiferromagnets, respectively, on top of their uniform states. The corresponding spin-current density (polarized along the $z$ axis) is given by $j^s_i = - A \hat{\mathbf{z}} \cdot (\mathbf{n} \times \partial_i \mathbf{n}) \approx - A \partial_i \phi$ both for ferromagnets and antiferromagnets, which can be obtained from the spin continuity equation~\cite{HalperinPR1969}. A spin current is carried by a gradient of the angle $\phi$ while traversing the ground-state manifold, which is analogous to a supercurrent in conventional superfluids that is carried by a finite gradient of the wavefunction phase, and thus it is referred to as a spin supercurrent~\cite{SoninAP2010}. Differing from spin-wave modes accounting for small fluctuations of the order parameter, a spin supercurrent can be carried by the large nonperturbative variation.

\begin{figure}
\includegraphics[width=0.9 \columnwidth]{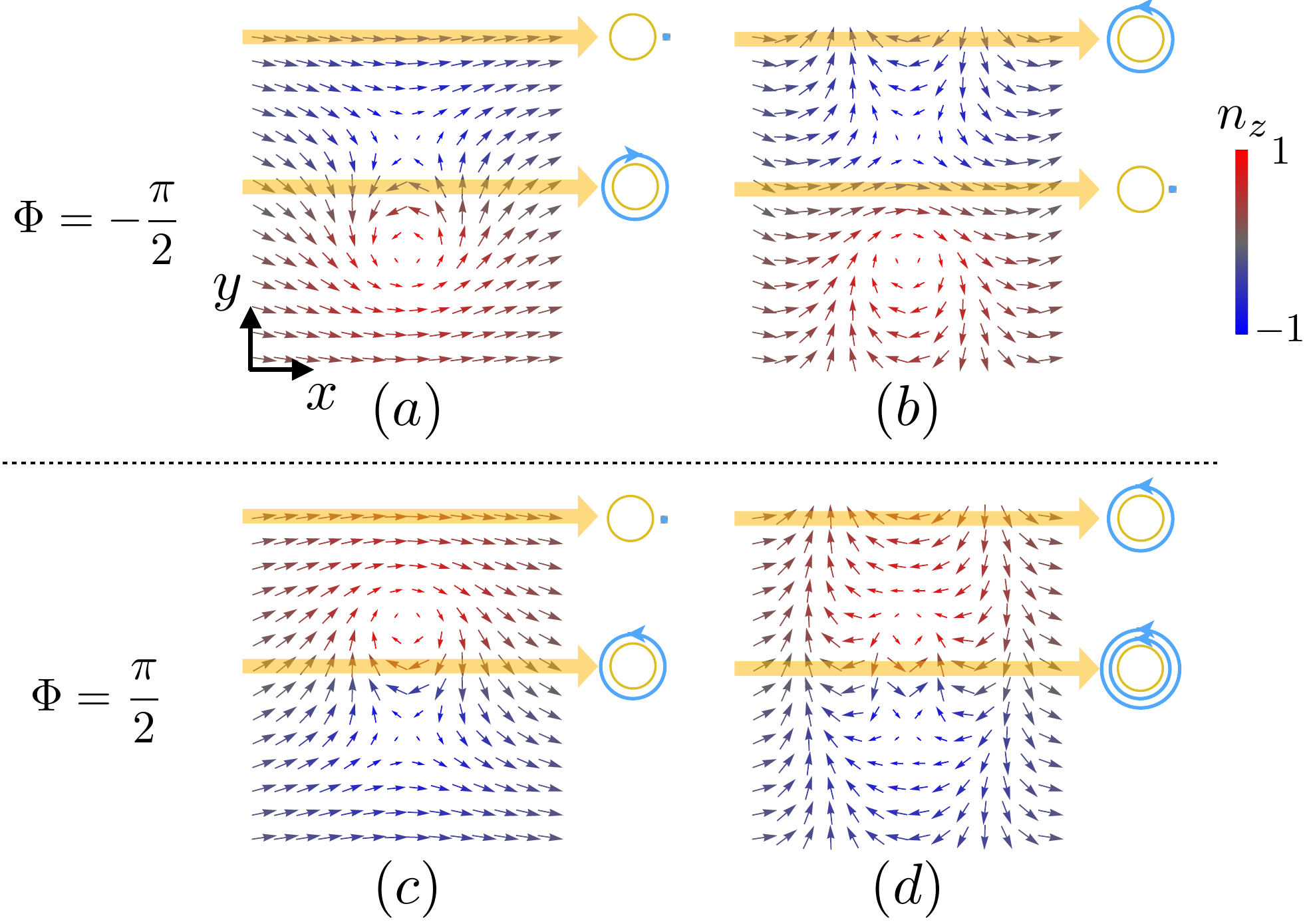}
\caption{A bimeron skyrmion [Eq.~(\ref{eq:ansatz})] with $Q = 1$, angle $\Phi$ and azimuthal angle background $\phi = k x$: (a) $\Phi = - \pi/2 \, , k = 0$, (b) $\Phi = - \pi/2 \, , k = 2 \pi / L_x$, (c) $\Phi = \pi/2 \, , k = 0$, (d) $\Phi = \pi/2 \, , k = 2 \pi / L_x$, where $L_x$ is the sample length in the $x$ direction. The winding numbers of the azimuthal angle across certain horizontal lines are shown on the right of plots. The background spin current $\propto k$ induces an energy difference between two configurations $\Phi = \pm \pi/2$, which otherwise possess the same energy.}
\label{fig:fig2}
\end{figure}

We study the interaction of a skyrmion with a spin supercurrent within linear response, by considering the skyrmion as a perturbation to the otherwise uniform spin-current background. We assume that the skyrmion texture except for its zero-energy mode $\phi$ is rigid enough to neglect the effect of its change on the interaction between the skyrmion and the spin current. By plugging $\mathbf{n}$ [Eq.~(\ref{eq:ansatz})] with $\phi = \mathbf{k} \cdot \mathbf{r} = k (x \cos \phi_k + y \sin \phi_k)$ to the Hamiltonian $U$ [Eq.~(\ref{eq:U})], we obtain our first main result, the skyrmion energy to linear order in $k$:
\begin{equation}
U(\Phi; \mathbf{k}) = U_0 + A R_1 |\mathbf{k}| \sin (\Phi - \phi_k) \, ,
\end{equation}
where $U_0$ is the energy with $k = 0$, $R_1 \equiv I_1 R_0 + I_2 B / (A R_0)$ represents the effective size with which the skyrmion interacts with the spin current, and $I_1 \approx 7.6$ and $I_2 \approx 54$ are numerical constants. The explicit expressions for $I_1$ and $I_2$ are given by $I_1 = \pi^2 - (\pi / 2) \mathrm{Si}(2 \pi) \approx 7.6$ and $I_2 =\pi^2 \int_0^\infty dz \, [ (2/z - 1) (1 - \cos (2 \pi e^{-z})) e^{-z} - \pi e^{- 2z} \sin(2 \pi e^{-z}) + 2 z \pi^2 e^{- 3z} ] \approx 54$. The energy minimum is achieved when the angle is $\Phi = \phi_k - \pi / 2$, which can be understood as follows. The spin supercurrent $\propto k$ exerts opposite transverse forces on a vortex and an antivortex~\cite{NikiforovJETP1983}, which is analogous to the Magnus force in superconductors by which a charge supercurrent pushes a vortex and an antivortex in the opposite transverse directions~\cite{*[][{, and references therein.}] HalperinIJMPB2010}. The opposite forces on constituent vortices exert a torque on the skyrmion and thereby create the derived interaction term. Figure~\ref{fig:fig2} shows spin configurations for several cases. The winding numbers of the azimuthal angle across two horizontal lines are shown on the right. Since the larger winding number costs the higher exchange energy, the winding number can serve as good indicators for the energy. In the absence of the spin current $k = 0$, the energies for the two angles $\Phi = \pm \pi/2$ are equal. However, in the presence of the spin current, $k > 0$, $\Phi = \pi /2$ case has a higher energy than $\Phi = - \pi/2$ case, as can be seen from their winding numbers.

\subsection{Spin-supercurrent-induced skymion motion}
\label{sec:3-1}
\begin{figure}
\includegraphics[width=0.8\columnwidth]{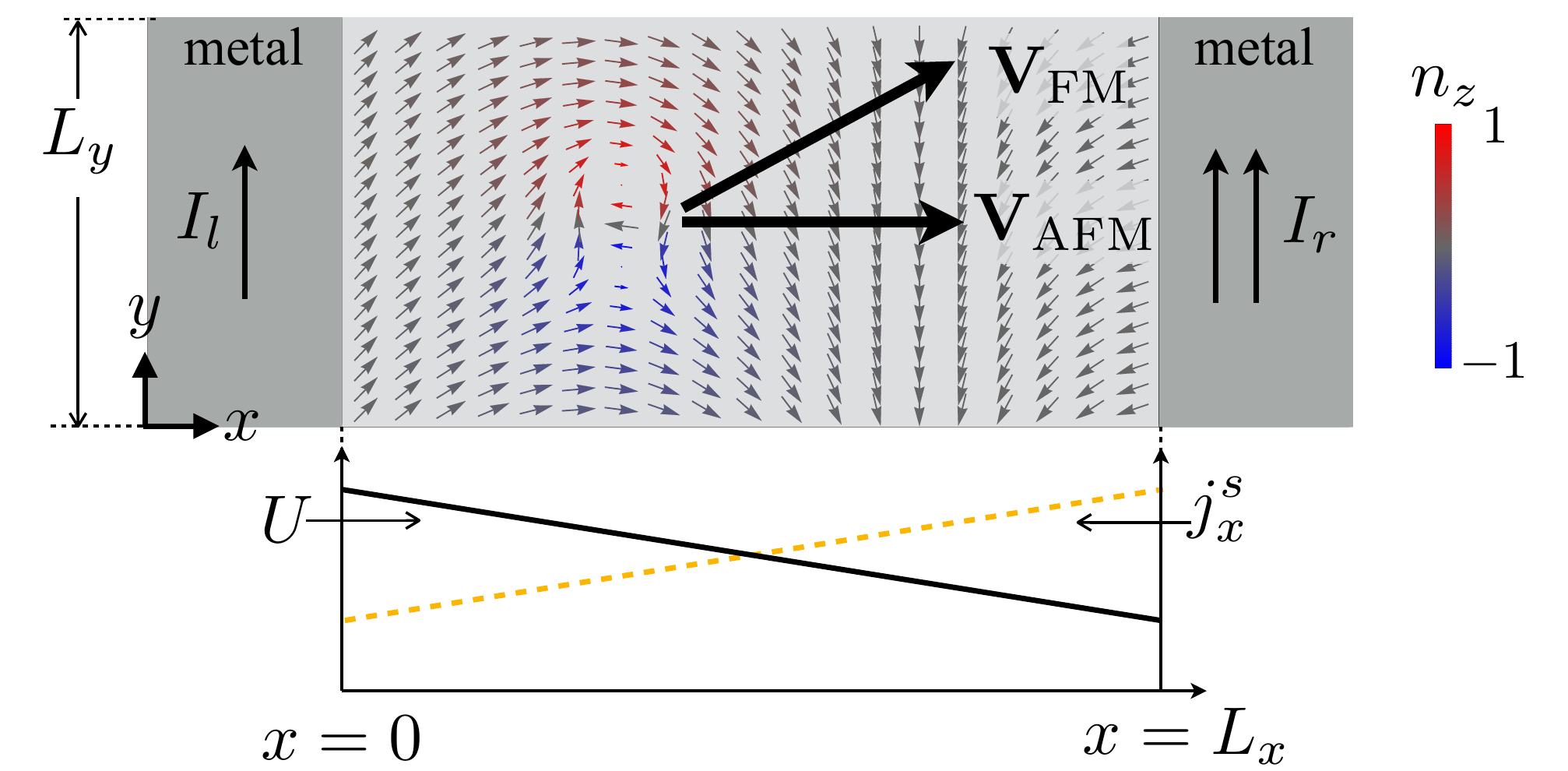}
\caption{A bimeron skyrmion in the background of a nonuniform spin current $j^s_x$ that is induced via the interfacial spin Hall effects by asymmetric charge currents $I_l$ and $I_r$ in the left and the right proximate metals, respectively. The skyrmion experiences a force to the right, where the spin current is maximum. In ferromagnets, the steady-state velocity $\mathbf{V}_\text{FM}$ has a transverse component to the force due to the skyrmion Hall effect. In antiferromagnets, $\mathbf{V}_\text{AFM}$ is parallel to the force.}
\label{fig:fig3}
\end{figure}

Below, we show that a skyrmion can be driven by a nonuniform spin supercurrent. For the spin supercurrent that varies much slowly in space compared to the skyrmion size, the skyrmion will be in local equilibrium by adjusting its angle perpendicular to the local spin current (i.e., $\Phi = \phi_k (X, Y) - \pi / 2$) such that the energy is given by
\begin{equation}
\label{eq:UXY}
U(X, Y) = U_0 - A R_1 |\boldsymbol{\nabla} \phi| (X, Y) \, .
\end{equation}
The spin supercurrent in a magnet can be induced by attaching heavy metals such as Pt to its boundaries and subsequently using the interfacial spin Hall effect as shown in Ref.~\cite{TakeiPRL2014}. See Fig.~\ref{fig:fig3} for illustrations. For the charge currents $I_l$ and $I_r$ in the left and the right metals, the boundary conditions for the spin current are given by
\begin{equation}
\label{eq:boundary}
\begin{split}
j^s_x &= \vartheta I_l - \gamma \dot{\phi} \, , \quad \text{for } x = 0 \, , \\
j^s_x &= \vartheta I_r + \gamma \dot{\phi} \, , \quad \text{for } x = L_x \, ,
\end{split}
\end{equation}
in linear response~\cite{TserkovnyakPRB2014}. The left-hand sides are the spin currents in the magnet at its interface with the metals. The first terms on the right-hand sides are spin torques exerted by the charge currents via the spin Hall effects. The second terms are spin pumping from the magnet into the metals. Here, $\vartheta$ is the coefficient parametrizing the dampinglike torque on the magnet induced by the charge current, which is related to the effective interfacial spin Hall angle $\Theta$ by $\vartheta = \hbar \tan \Theta / 2 e d_y$ with $d_y$ the thickness of the metals along the $y$ direction, $-e$ the charge of electrons; $\gamma \equiv \hbar g^{\uparrow \downarrow} d_z / 4 \pi$ is the parameter for the spin pumping at the interface with $g^{\uparrow \downarrow}$ the effective interfacial spin-mixing conductance and $d_z$ the thickness along the $z$ direction. For the top and bottom boundaries, we consider an open boundary condition $\mathbf{j}_s \cdot \hat{\mathbf{y}} \equiv 0$~\cite{Brown1963, HoffmanPRB2013, ReitzPRB2018}. The spin continuity equation in the bulk is given by 
\begin{equation}
\dot{s}_z + \boldsymbol{\nabla} \cdot \mathbf{j}^s + \alpha s \dot{\phi} = 0 \ ,
\label{eq:bulk}
\end{equation}
both for ferromagnets~\cite{TakeiPRL2014} and antiferromagnets~\cite{TakeiPRB2014}, where $s_z$ is the $z$ component of the spin density, $\alpha$ is a dimensionless number parametrizing damping referred to as the Gilbert constant~\cite{GilbertIEEE2004}, and $s$ is the saturated spin density. By solving the bulk equation of motion subjected to the boundary conditions with uniform $I_l$ and $I_r$, the stead-state solution can be obtained: 
\begin{equation}
\phi(x, t) = \omega t - [(\vartheta I_l - \gamma \omega) / A] x + (\alpha s \omega / 2 A) x^2 + \phi_0 \, ,
\end{equation}
where $\phi_0$ is arbitrary and the frequency is given by~\cite{TakeiPRL2014, TakeiPRB2014}
\begin{equation}
\omega = \vartheta (I_l - I_r) / (2 \gamma + \alpha s L) \, .
\end{equation}

The induced spin current is non-uniform in space and thus creates a nontrivial energy landscape for the skyrmion. When $I_r > I_l \geq 0$ with $\vartheta > 0$, from $U(X, Y)$ [Eq.~(\ref{eq:UXY})] with the aforementioned solution for $\phi$, we obtain the second main result, the force on the skyrmion:
\begin{equation}
F_X = - \frac{d U}{d X} = R_1 \frac{\alpha s \vartheta (I_r - I_l)}{2 \gamma + \alpha s L}  \, ,
\label{eq:FX}
\end{equation}
and $F_Y = 0$. The direction of the resultant motion depends on the nature of the magnet. For ferromagnets, the equations of motion for $X$ and $Y$ are given by~\cite{ThielePRL1973, TretiakovPRL2008, KimPRB2017}
\begin{equation}
\label{eq:eom}
\begin{split}
4 \pi s Q \dot{Y} + \alpha s I_A \dot{X} &= F_X \, , \\
- 4 \pi s Q \dot{X} + \alpha s I_A \dot{Y} &= F_Y \, ,
\end{split}
\end{equation}
where $I_A \approx \pi \int_0^\infty dz \, z [ z^{-2} \sin^2 \zeta_0 (z) + \left\{ \zeta'_0 (z) \right\}^2 ] \approx 13$ is a numerical constant. For the present case, $F_Y = 0$ and thus the solution is given by
\begin{eqnarray}
\dot{X}_\text{FM} &=& \frac{\alpha s I_A}{(4 \pi s)^2 + (\alpha s I_A)^2} F_X \, , \\
\dot{Y}_\text{FM} &=& \frac{4 \pi s Q}{(4 \pi s)^2 + (\alpha s I_A)^2} F_X \, .
\end{eqnarray}
Note that the motion is deflected from the direction of the force, exhibiting the skyrmion Hall effect~\cite{LitzuisNP2016, JiangNP2017}. For antiferromagnets, the equations of motion for $X$ and $Y$ are given by~\cite{TvetenPRL2013, BarkerPRL2016}:
\begin{equation}
\begin{split}
M \ddot{X} + \alpha s I_A \dot{X} &= F_X \, , \\
M \ddot{Y} + \alpha s I_A \dot{Y} &= F_Y \, ,
\end{split}
\end{equation}
where $M$ is the inertial mass for skyrmions~\cite{TvetenPRL2013, BarkerPRL2016}. The steady-state velocity for our current situation is given by
\begin{equation}
\label{eq:vafm}
\dot{X}_\text{AFM} = \frac{F_X}{\alpha s I_A} \, , \quad \dot{Y}_\text{AFM} = 0 \, .
\end{equation}
The skyrmion motion is driven by a dissipation-induced gradient of a spin current, not by a spin current per se, as manifested in the $\alpha$-dependence of the force $F_X$ [Eq.~(\ref{eq:FX})]. See Fig.~\ref{fig:fig3} for the spin texture $\mathbf{n}(x, y)$ [Eq.~(\ref{eq:ansatz})] when $I_r = 2 I_l > 0$ and $\gamma = 0$. For numerical estimates, we take the following material parameters: lattice constant $a \sim 0.5$ nm, $\sqrt{A/K} \sim 2 a$, $B \sim A a^2$, $s = \hbar / a^2$~\cite{RegnaultJPF1982, YoshiyamaJPC1984}, and the damping parameter $\alpha = 0.1$, which results in $R_0 \sim 1.5$ nm and $R_1 \sim 20$ nm. We also take $L_x = 100$ nm, $d_y = 5$ nm and $d_z = 10$ nm for sample geometry and $\Theta = 0.1$ (obtained for Pt\textbar permally interface~\cite{LiuPRL2011}) for the interfacial spin Hall effects. When the applied current density is $I_r / (d_y d_z) = 10^{10}$ A/m$^2$, we obtain $V_\text{FM} \sim 0.1$ m/s and $V_\text{AFM} \sim 1$ m/s.

\subsection{Skyrmion racetrack}
\label{sec:3-2}
Going beyond from the previous case for a uniform force, where the uniform charge currents are considered. let us now allow the boundary charge currents to be inhomogeneous, $I_l (y)$ and $I_r (y)$. We seek a steady-state solution with $\dot{s}_z = 0$ and $\dot{\phi} \equiv \omega$. Then finding the angle configuration for given boundary charge currents constitutes the problem of solving Poisson's equation $\boldsymbol{\nabla}^2 \phi = \alpha s \omega / A$ [Eq.~(\ref{eq:bulk})] with the Neumann boundary conditions [Eq.~(\ref{eq:boundary}) and open boundary conditions $j^s_y = 0$ at $y = 0$ and $y = L_y$, where $L_y$ is the sample length in the $y$ direction], which is known to have one and the only one solution upto a constant~\cite{Jackson}. The frequency $\omega$ can be obtained by the bulk-boundary compatibility condition of Poisson's equation: 
\begin{equation}
\int_V \, \boldsymbol{\nabla}^2 \phi = \oint_{\partial V} \hat{\boldsymbol{\nu}} \cdot \boldsymbol{\nabla} \phi \, ,
\end{equation}
with $\hat{\boldsymbol{\nu}}$ the outward normal vector to the boundary, which, in our case, corresponds to
\begin{equation}
\alpha s \omega L_x L_y = \vartheta \int dy \, (I_l (y) - I_r(y)) - 2 \gamma \omega L_y \, .
\end{equation}
This expresses the conservation of spin: The left-hand side is the spin-dissipation rate in the bulk, the first term on the right-hand side is the rate of the current-induced spin injection, and the second term is the spin pumping from the magnet to the metals. The solution is given by
\begin{equation}
\omega = \frac{\vartheta}{2 \gamma + \alpha s L_x} \frac{1}{L_y} \int dy \, (I_l (y) - I_r(y)) \, ,
\end{equation}
which generalizes the previous results for uniform currents~\cite{TakeiPRL2014, TakeiPRB2014}. Note that the differential equations and the compatibility condition are all linear in the currents $I_l(y)$ and $I_r(y)$. Therefore, any linear superposition of two solutions $\phi = \phi_1 + \phi_2$, where $\phi_1$ and $\phi_2$ are the solutions for the different charge-current pairs, is also a solution to the problem for the added charge currents.

\begin{figure}
\includegraphics[width=\columnwidth]{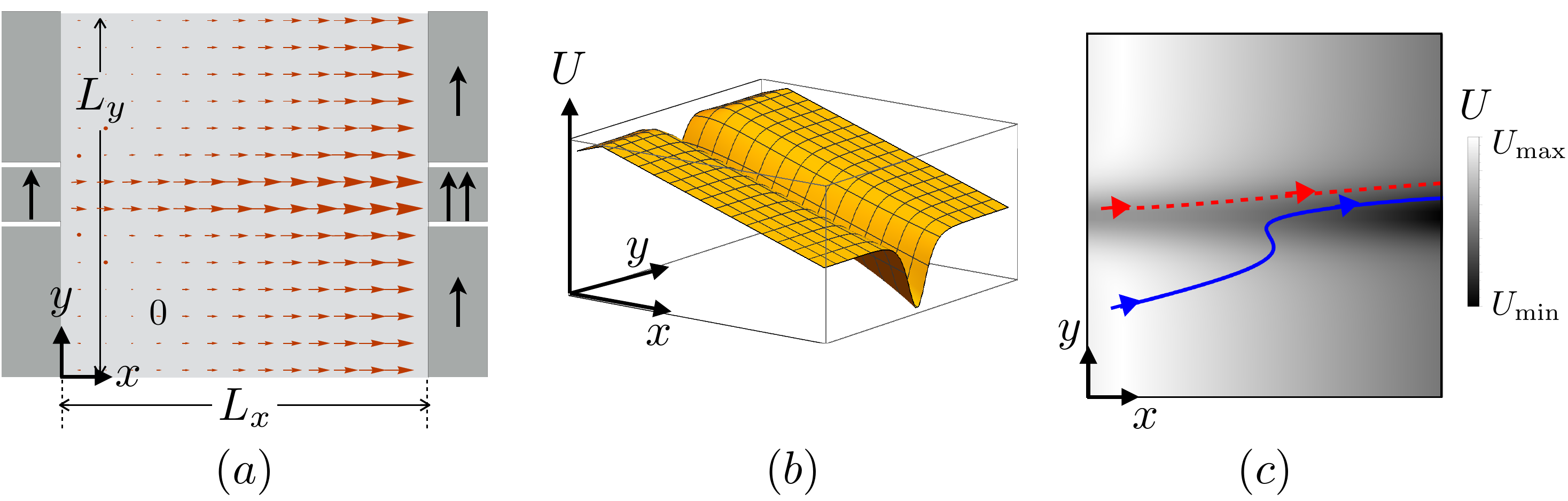}
\caption{(a) The spin current (shown by the red arrows) induced by nonuniform charge currents (shown by the black arrows). (b) The corresponding energy landscape for a skyrmion. (c) Two trajectories of a ferromagnetic skyrmion.}
\label{fig:fig4}
\end{figure}

A racetrack for a skyrmion can be engineered as follows. First, a skyrmion can be localized along the vertical center of the sample by injecting a large spin current only near the center so that the spin current flows dominantly along the line defined by $y = L_y/2$. This decreases the skyrmion energy along the line, engendering a racetrack. Then, a skyrmion can be driven to the right along the racetrack by inducing an additional spin current on the right only. As an example, we consider the case where the left and the right charge currents are given by $\vartheta I_l (y) = 2 \sech((y - L_y/2) / d)$ and $\vartheta I_r (y) = 2 \sech((y - L_y/2) / d) + 2.5$, respectively, with sample geometry $L_x = 50$ and $L_y = 500$, and effective racetrack width $d = 20$. Here, we measure energy, length, and time in $A$, $R_0$, and $s R_0^2 / A$, respectively. Figure~\ref{fig:fig4}(a) shows the spin current $\mathbf{j}^s$ obtained by solving Poisson's equation with the given Neumann boundary conditions. Note that the magnitude of the spin current is the largest along the racetrack. The energy landscape is shown in Fig.~\ref{fig:fig4}(b). Figure~\ref{fig:fig4}(c) shows two trajectories of a ferromagnetic skyrmion which are obtained by solving the equations of motion~(\ref{eq:eom}) for the given energy landscape with $\alpha = 0.4$ and $R_1 = 10$. Both trajectories are focused on the racetrack. The sudden velocity change of the solid trajectory is caused by the vertical force localized near the racetrack that pushes the skyrmion in the negative $x$ direction via the skyrmion Hall effect momentarily. We would like to remark here that Fig.~\ref{fig:fig4}(a) illustrates one way to inject an inhomogeneous spin current via a series of nonmagnetic metals. It should be also possible, in principle, to use a single nonmagnetic metal with engineered spatial variation of material properties such as the spin Hall angle that is directly related to spin-current injection.

\section{Discussion}
\label{sec:4}
We have shown that a bimeron skyrmion in easy-plane magnets can be driven by engineering its energy landscape via its interaction with a spin supercurrent. The induced dynamics can be controlled non-locally by charge currents in proximate normal metals. One assumption of our theory is the presence of the perfect U(1) spin-rotational symmetry of the system, which yields the conservation of spin at the Hamiltonian level and thereby eases the theoretical treatment of a spin current. However, even when the U(1) symmetry is weakly broken, e.g., by additional anisotropy within the easy plane, a superfluid-like spin current can be induced by applying sufficiently large currents on metals~\cite{SoninAP2010, QaiumzadehPRL2017, HillPRL2018} and thus we expect that our theory for the skyrmion motion is generically applicable. In addition, we would like to mention that including dipolar interactions that can affect both the statics and the dynamics of a bimeron skyrmion can be an important research topic for utilizing bimeron skyrmions, but it is beyond the scope of our current work. Lastly, we envision that the main idea of the present work, to control a skyrmion by manipulating a background spin texture within its ground-state manifold, can be extended to general cases even beyond magnetism, wherever a localized soliton exists in a controllable background with the low-energy manifold.

\begin{acknowledgments}
We acknowledge useful discussions with Héctor Ochoa and Yaroslav Tserkovnyak. This work was supported by the startup fund at the University of Missouri.
\end{acknowledgments}

\appendix

\section{The spin continuity equation for easy-plane ferromagnets in their uniform ground states}
\label{app:fm}

Here, we derive the continuity equation for spin (polarized along the $z$ axis) for easy-plane ferromagnets on top of uniform ground states. For ferromagnets, the long-wavelength low-energy dynamics can be described by the following Hamiltonian:
\begin{equation}
U = \frac{1}{2} \int dxdy \left[ A (\nabla \phi)^2 + K n_z^2 \right] \, , 
\end{equation}
where $\mathbf{n} = (\sqrt{1 - n_z^2} \cos \phi, \sqrt{1 - n_z^2} \sin \phi, n_z)$ is the unit vector in the direction of the local spin density and $A$ and $K$ are positive exchange and anisotropy coefficients, respectively. For long-wavelength dynamics, the last term $\propto B$ in Eq.~(\ref{eq:U}), which is quartic in the spatial gradient, can be neglected over the term $\propto A$, which is quadratic in the spatial gradient. The spin density (polarized along the $z$ direction) is given by $s_z = s n_z$, where $s$ is the saturated spin density. For the low-energy theory, the two dynamic variables are $s_z$ and $\phi$. As we learn from quantum mechanics, the spin angular momentum density $s_z$ is the generator of the spin rotations about the $z$ axis~\cite{LL3}. In other words, the two variables form a pair of canonical conjugate variables, satisfying the Poisson bracket $\{ \phi(\mathbf{r}, t), s_z (\mathbf{r}', t) \} = \delta (\mathbf{r} - \mathbf{r}')$~\cite{Goldstein}. The corresponding equations of motion can be derived from the Hamiltonian and the Poisson bracket:
\begin{eqnarray}
\dot{s}_z &=& \{ s_z, U \} = A {\boldsymbol{\nabla}}^2 \phi \, , \\
\dot{\phi} &=& \{ \phi, U \} = \frac{K s_z}{s^2} \, .
\end{eqnarray}
The first equation is the spin continuity equation and the second equation is the spin Josephson relation. From the right-hand side of the continuity equation, we can identify the spin current as $\mathbf{j}^s = - A \boldsymbol{\nabla} \phi$. The spin current is carried by the spatial gradient of the azimuthal angle that represents the spontaneous breaking of the U(1) spin-rotational symmetry of the Hamiltonian, and it is analogous to the charge current in superconductors that is carried by the spatial gradient of the wavefunction phase that represents the spontaneously broken U(1) phase symmetry. For this analogy, this spin current in easy-plane ferromagnets is referred to as a spin supercurrent~\cite{SoninAP2010}. The damping term, which makes a spin supercurrent different from a charge supercurrent, can be accounted for by considering the Rayleigh dissipation function $R = \alpha s \int dxdy \, \dot{\mathbf{n}}^2 / 2$, which is the half of the energy dissipation rate through the magnetic dynamics. For the low-energy dynamics, $R \approx \alpha s \int dxdy ( \dot{\phi}^2 + \dot{s}_z^2 / s^2 ) / 2$. The equations of motion with the damping are then given by
\begin{eqnarray}
\dot{s}_z &=& \{ s_z, U \} - \frac{\delta R}{\delta \dot{\phi}} = A {\boldsymbol{\nabla}}^2 \phi - \alpha s \dot{\phi} \, , \\
\dot{\phi} &=& \{ \phi, U \} + \frac{\delta R}{\delta \dot{s}_z} = \frac{K s_z}{s^2} + \frac{\alpha \dot{s}_z}{s} \, .
\end{eqnarray}

\section{The spin continuity equation for easy-plane antiferromagnets in their uniform ground states}
\label{app:afm}

Here, we derive the continuity equation for spin (polarized along the $z$ axis) for easy-plane antiferromagnets on top of their uniform states. For antiferromagnets, the long-wavelength low-energy dynamics can be described by the following Hamiltonian:
\begin{equation}
U = \frac{1}{2} \int dxdy \left[ A (\nabla \phi)^2 + K n_z^2 + \mathbf{s}^2 / \chi \right] \, , 
\end{equation}
where $\mathbf{n} = (\sqrt{1 - n_z^2} \cos \phi, \sqrt{1 - n_z^2} \sin \phi, n_z)$ is the unit vector in the direction of the local N{\'e}el order, $A$ and $K$ are positive exchange and anisotropy coefficients, respectively, $\chi$ represents the magnetic susceptibility, and $\mathbf{s} = (s_x, s_y, s_z)$ is the spin density. Note that the last term $\propto s_z^2$ is not included in Eq.~(\ref{eq:U}) which is the energy functional for the static configuration of the magnet, since $s_z$ is strongly suppressed and thus can be neglected for the static case. However, it is necessary to be included in the Hamiltonian to describe the dynamics of the antiferromagnets. Our interest is in the dynamics of the azimuthal angle $\phi$ and the spin density $s_z$ along the $z$ axis, and, as we discussed above for a ferromagnetic case, they form a pair of canonical conjugate variables, satisfying the Poisson bracket $\{ \phi(\mathbf{r}, t), s_z (\mathbf{r}', t) \} = \delta (\mathbf{r} - \mathbf{r}')$~\cite{Goldstein}. Correspondingly, the dynamics of these two variables $\phi$ and $s_z$ are decoupled from the dynamics of the other variables $n_z, s_x$, and $s_y$.The equations of motion for $\phi$ and $s_z$ can be derived from the Hamiltonian and the Poisson bracket:
\begin{eqnarray}
\dot{s}_z &=& \{ s_z, U \} = A {\boldsymbol{\nabla}}^2 \phi \, , \\
\dot{\phi} &=& \{ \phi, U \} = \frac{s_z}{\chi} \, .
\end{eqnarray}
After including the damping through the Rayleigh dissipation function $R = \alpha s \int dxdy \, (\dot{\phi}^2 + \dot{s}_z^2 / s^2) / 2$, where $s$ is the saturated spin density, we obtain 
\begin{eqnarray}
\dot{s}_z &=& \{ s_z, U \} - \frac{\delta R}{\delta \dot{\phi}} = A {\boldsymbol{\nabla}}^2 \phi - \alpha s \dot{\phi} \, , \\
\dot{\phi} &=& \{ \phi, U \} + \frac{\delta R}{\delta \dot{s}_z} = \frac{s_z}{\chi} + \frac{\alpha \dot{s}_z}{s} \, .
\end{eqnarray}
Note the identical forms of the two equations, the spin continuity equation and the spin Josephson relation, between ferromagnet and antiferromagnet cases.

\bibliographystyle{/Users/kimsek/Dropbox/School/Research/apsrev4-1-with-title}
\bibliography{/Users/kimsek/Dropbox/School/Research/master}

\end{document}